\documentclass[review]{elsarticle}
\usepackage{graphicx}
\usepackage{bm}
\usepackage{lineno,hyperref}
\modulolinenumbers[5]

\journal{Journal of High Energy Astrophysics}

\def\beq{\begin{eqnarray}}
\def\eeq{\end{eqnarray}}

\begin{document}

\begin{frontmatter}

\title{A possible feedback mechanism of outflows from a black hole hyperaccretion disk in the center of jet-driven iPTF14hls}

\author{Tong Liu, Cui-Ying Song, Tuan Yi, Wei-Min Gu}
\address{Department of Astronomy, Xiamen University, Xiamen, Fujian 361005, China; tongliu@xmu.edu.cn}

\author{Xiao-Feng Wang}
\address{Department of Physics and Tsinghua Center for Astrophysics, Tsinghua University, Beijing 100084, China}

\begin{abstract}
iPTF14hls is an unusually bright, long-lived II-P supernova (SN), whose light curve has at least five peaks. We propose that the outflows from the black hole hyperaccretion systems in the center of the collapsars should continuously inject into the envelope. For a jet-driven core-collapsar model, the outflow feedback results in prolonging the accretion timescale and fluctuating accretion rates in our analytic solutions. Thus, the long period of luminous, varying SN iPTF14hls might originate from the choked jets, which are regulated by the feedback of the strong disk outflows in a massive core-collapsar. One can expect that jet-driven iPTF14hls may last no more than approximately 3,000 days, and the luminosity may quickly decrease in the later stages. Moreover, the double-peak light curves in some SNe might be explained by the outflow feedback mechanism.
\end{abstract}


\end{frontmatter}


\section{Introduction}
	
A peculiar supernova (SN), iPTF14hls, was discovered by the Intermediate Palomar Transient Factory (iPTF) wide-field camera survey in September 2014 \citep{Arcavi2017}. Its redshift is $z$=0.0344, and the total energy in radiation is approximately $2.2 \times 10^{50} ~\rm erg$. According to the observations on the broad Balmer series P Cygni lines, iPTF14hls was identified as a Type II-P SN, but the evolution of the light curve is much slower than that of a typical SN. In particular, this SN has at least five peaks during which the luminosity varied by approximately 50$\%$ in the light curve lasting over 600 days. The velocities of the Fe$_{\rm II}$ and H$_\alpha$ lines remain at approximately 4,000 and 8,000 $\rm km ~s^{-1}$, respectively. A hydrogen-rich and massive star has been suggested as the progenitor of iPTF14hls \citep{Arcavi2017,Milisavljevic2018}. Furthermore, \citet{Yuan2018} reported the suspected association between this SN and a gamma-ray source detected by \emph{Fermi}-LAT.
	
The spin-down magnetar \citep[e.g.,][]{Kasen2010,Woosley2010} and the fallback accreting black hole \citep[BH, e.g.,][]{Dexter2013,Liu2013,Liu2017,Liu2018} are the standardized central engine models in the core collapsars of the power observable gamma-ray bursts (GRBs) associated with SNe or isolated SNe. \citet{Arcavi2017} noticed a $t^{-5/3}$ decline rate after day 450, which might support the BH accretion model better than magnetar power \citep[also see][]{Sollerman2018}. \citet{Arcavi2017} suggested that instabilities in the accretion disk could produce the variability. Several competitive and plausible models were presented. \citet{Dessart2018} insisted that a magnetar born in a blue-supergiant star could explain this SN. \citet{Wang2018} studied the process of the fallback accretion onto a neutron star (NS) as the central engine of iPTF14hls. \citet{Soker2018} proposed that iPTF14hls can be powered by an exotic NS spiralling-in inside the envelope of a massive companion star. \citet{Andrews2018} reported a moderate-resolution spectrum of iPTF14hls after day 1153, which is evidence of the interaction between the dense circumstellar medium (CSM) and shock. Such features cannot be explained by the magnetar model \citep{Woosley2018}. Another mechanism, the pulsational pair-instability model, has also been discussed \citep{Woosley2017,Woosley2018}, which can interpret the suspected homologous outburst recorded by the Palomar Observatory Sky Survey (POSS) in February 1954. Moreover, \citet{Chugai2018} argued that there might have been an explosion of a massive star at day 450 before iPTF14hls because of the observations of the H$_\alpha$ lines.
	
We studied the BH inflow-outflow hyperaccretion systems of the different progenitors of GRBs, i.e., the mergers of compact objects \citep[BH-NS or NS-NS, see e.g.,][]{Liu2017,Song2018} and massive collapsars \citep[e.g.,][]{Liu2017,Liu2018,Song2019}. The characteristics of the progenitors can be constrained, and we found associated electromagnetic phenomena such as kilonovae and SN bumps ($^{56}$Ni bumps). SN bumps refer to the late-time optical bumps in the afterglows of GRBs, which mainly originate from the $^{56}$Ni decay and its daughter $^{56}$Co to $^{56}$Fe. In the accretion framework, the outflows play important roles in the production of kilonovae and SN bumps, which compete on the budgets of the disk masses and energies with the inflows to launch the jets to power GRBs. Consequently, the outflows are potentially the main element factories \citep[e.g.,][]{Liu2017,Song2019}.
	
In the core-collapsar scenario, the materials of the envelope fall towards the central compact object, which might be a BH, and are converted into an accretion disk. Meanwhile, the jets launch from the accretion system to break through the envelope, and the outflows from the disk have been injected into the envelope. There are interactions on the masses and energies between the progenitors and outflows in the collapsars. We consider that the feedback mechanism of the outflows from the disks might exist in stellar-scale collapsars.
	
In this paper, we present a simplified picture and only focus on the pre-SN model excluding the explosion process \citep[e.g.,][]{Heger2003,Woosley2012}; then, we investigate the feedback effects of the outflows from the BH hyperaccretion process on the mass supply to the stellar envelope. This mechanism may explain the timescale and light curve of iPTF14hls. This paper is organized as follows. In Section 2, we propose our toy model on the outflow feedback. Conclusions and discussion are included in Section 3.
	
\section{Outflows feedback model}
	
\begin{figure}
\centering
\includegraphics[angle=0,scale=0.35]{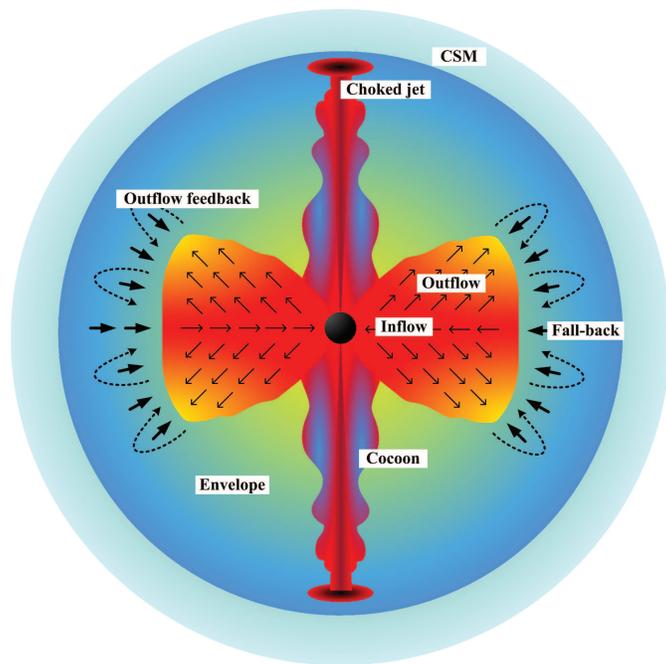}
\caption{Schematic picture of choked jets and strong outflows from a BH hyperaccretion system in a core collapsar scenario.}
\end{figure}
	
\subsection{Accretion timescale}
	
A stellar-mass BH might be born in the center after a massive progenitor collapses. The materials from the envelope fall back towards the BH to trigger the hyperaccretion process and launch the Blandford-Znajek \citep[BZ, see][]{Blandford1977} jets \citep[e.g.,][]{Heger2003,Liu2018}. Once the jets break out from the envelope and one aligns along the line of sight, an observable long- or ultra-long-duration GRB (LGRB or ULGRB) is formed. The fall-back hyperaccretion system has been widely proposed as the central engine for LGRB-SN, even for ULGRB-SN \citep[e.g.,][]{Wu2013,Gao2016,Liu2018,Song2019}. A long-lived hyperaccretion process and the modest thickness and density of the stellar envelope and CSM can make the successful jets. Thus, the low-metallicity and massive progenitor stars are considered to be suitable for LGRB-SN \citep[e.g.,][]{Liu2018,Song2019}.
	
Once the jets are choked in the stellar envelope or CSM, their energy has been injected into the envelope; then, only the thermalized radiation from the photosphere can be detected \citep[e.g.,][]{Maeda2003,Lazzati2012,He2018}. The CSM can make similar contributions with the envelope on thermalizing the radiation and bringing the high-velocity lines \citep[e.g.,][]{Andrews2018,Woosley2018,Luo2018}. The velocities could be maintained throughout this event since the jets can last sufficiently long as well as the accretion process. The jets drive the highly anisotropic explosions, which might be more energetic than the isotropic explosion \citep[e.g.,][]{Heger2003,Halevi2018,Hayakawa2018}. For a typical SN explosion, the average $^{56}$Ni mass is $0.4\pm0.2~M_\odot$ \citep{Cano2017}. \citet{Kann2016} derived more than 2 $M_\odot$ $^{56}$Ni mass for SN 2011kl associated with ULGRB GRB 111209A. Actually, 10$\%$ of the strong disk outflow materials converted into $^{56}$Ni in the center of the low-matellicity and massive progenitors are satisfied with the mean luminosity of all observed (U)LGRB-SN samples \citep{Song2019}. In our feedback model on iPTF14hls, we consider that the jets cannot break out from the envelope or CSM and that they inject the energy to the envelope or CSM to power it. In other words, iPTF14hls might be a jet-driven SN; then, we can ignore the discussion on the process of the isotropic SN explosion. The materials of the outflows from the disk can be injected into the envelope and recycled via accretion by the BH to launch the jets, which subsequently prolong the SN timescale and produce the light curve of iPTF14hls.
	
Here, we only focus on the outflow dynamics. One of the challenges is the accretion timescale. We propose that the massive outflows can solve this issue. The outflows from the disk can inject the mass and energy into the rest of the envelope of the progenitor as long as the envelope can block the escape of the outflows. The dynamic picture of iPTF14hls is shown in Figure 1.
	
Outflows from the BH accretion disks are the topics of an enormous amount of theoretical \citep[e.g.,][]{Narayan1994,Blandford1999,Liu2008,Liu2018,Gu2015}, simulated \citep[e.g.,][]{Stone1999,Ohsuga2005,Ohsuga2011,Jiang2014,Jiang2017,Yuan2014,Sadowski2015,Siegel2017} and observational \citep[e.g.,][]{Wang2013,Cheung2016,Parker2017} studies. The relationship between the accretion rate at the inner radius $\dot{M}_{\rm inner}$ and at the outer boundary $\dot{M}_{\rm outer}$ approaches a power-law with radius \citep[e.g.,][]{Yuan2014,Liu2018,Song2018}, and $\dot{M}_{\rm outer}$ roughly equals to the mass supply rate from the envelope $\dot{M}$. We refer to a parameter $f$ to describe the outflow rates, i.e.,
\beq
\dot{M}_{\rm o}=f  \dot{M},
\eeq
then $\dot{M}_{\rm inner}$ can be written as
\beq
\dot{M}_{\rm inner}=(1-f)  \dot{M}.
\eeq
For the super-Eddington accretion disk, $f$ should be larger than 0.9 \citep[e.g.,][]{Yuan2014,Liu2018,Song2018}.
	
The jet luminosity can be estimated by
\beq
L_{\rm j} = \eta_1 \dot{M}_{\rm inner} c^2.
\eeq
If we consider that the jets cannot break out from the envelope or CSM, which is thermalized to power a jet-driven SN, then there is an efficiency $\eta_2$ between the jet luminosity and the jet-driven SN luminosity.
The SN luminosity can be written as
\beq
L_{\rm SN} = \eta_1 \eta_2 \dot{M}_{\rm inner} c^2 = \eta \dot{M}_{\rm inner} c^2.
\eeq
From this equation, the SN luminosity is connected with the disk outflows. The typical value of $\eta$ can be set as approximately $10^{-5}$ \citep[e.g.,][]{Suwa2011,Nakauchi2013}.
	
Neglecting the mass and energy of outflows injected into the envelope, the mass supply rate can be concluded by \citep[e.g.,][]{Woosley2012,Liu2018}
\beq
\dot{M}=\frac{d M_{R}}{d t_{f}},
\eeq
where $M_{R}$ is the mass distribution function of the envelope, and $t_{f}$ is generally set as the free-fall timescale \citep[e.g.,][]{Woosley2012,Liu2018}.
	
If the mass and energy of outflows are assumed as the isotropic distribution, and $M_R$ mixing of the injection of outflows occurs, we can revisit $t_f$ using the similar computing method of the free-fall timescale. In fact, using the specific forces from the outflows, we obtain $\dot{M}_{\rm o}v_{\textrm{o}}/M_R$, where $v_{\textrm{o}}$ is the outflow velocities from the disk. These forces are usually too weak to offset the gravity, so $t_f$ still approaches the free-fall timescale, and the $t^{-5/3}$ decline is possibly retained in the light curve. Even so, the outflows change the mass distribution of the envelope. Moreover, the value of the outflow momenta are too small to enhance the strength of the explosion, but the decay of the radioactive elements produced in the outflows might partly contribute to the SN luminosity \citep[e.g.,][]{Surman2006,Song2019}. Here, we consider that the disk outflows mainly influence the jets, and the jets collide with the envelope or CSM to power iPTF14hls.
	
More importantly, the outflows are not immediately converted into fallback materials (see Section 2.3). In the case of the constant mass supply rate, we set $f \sim 0.9$, which means that $10\%$ of the disk mass is falling into the BH and $90\%$ of the disk mass is returned into the envelope to resupply the accretion disk in the endless cycle.
	
\begin{figure}
\centering
\includegraphics[angle=0,scale=0.6]{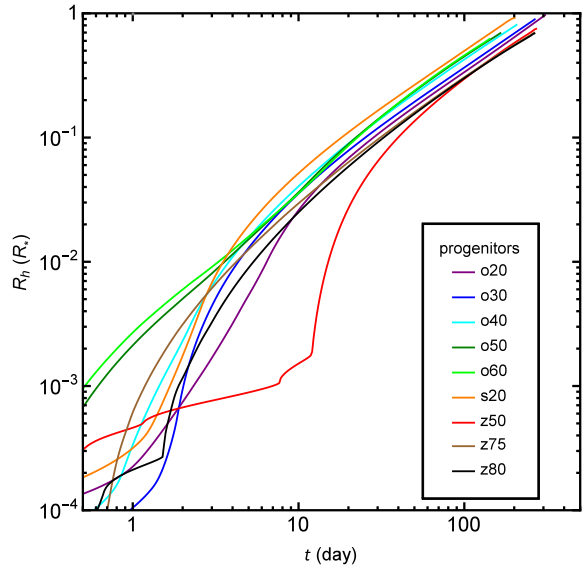}
\caption{Long-lived collapsars with choked jets excluding the outflow feedback with $f=0.9$, $\eta=10^{-5}$, and $\theta_{\rm j}=10^\circ$. The mean jet luminosity is satisfied with iPTF14hls.}
\end{figure}
	
\subsection{Choked jets}
	
In the collapsar scenario, we can obtain the radius of the jet head, which is the distance from the center to the jet head \citep[e.g.,][]{Matzner2003},
\beq
R_{\rm h}=\int_{t} c \beta_{\rm h} dt,
\eeq
where
\beq
\beta_{\rm h} =\frac{1}{1+\tilde{L}^{-1/2}},
\eeq
indicates the velocity in units of light speed of the jet head \citep[e.g.,][]{Matzner2003,Liu2018}, and
\beq
\tilde{L}=\frac{L_{\rm j} (t-R_{\rm h}/c)}{\pi \theta^2_{\rm j} R^2_{\rm h} \rho(R_{\rm h})c^3},
\eeq
where $\theta_{\rm j}$ and $\rho$ are the half opening angle of the jets and the radial density of the progenitor, respectively. If $R_{\rm h}$ is less than the radius of the progenitor $R_* \sim 10^{14}~\rm cm$ at the end of the accretion, the jets cannot break out from the envelope, which are called choked jets, as shown in Figure 1.
	
Figure 2 shows that the long-lived collapsars accompanied by the choked jets without consideration of the outflow feedback. The results in Figure 2 were calculated by Equations (1-8) as well as the most previous studies on LGRB or ULGRB progenitor models \citep[e.g.,][]{Heger2003,Quataert2012,Nakauchi2013,Ioka2016,Liu2018}. First, the mass supply rate was calculated by Equation (5), with the density profiles of the progenitor stars as shown in Figure 3 of \citet{Liu2018}. Second, the accretion rate at the inner radius could be derived from Equation (2). Third, the SN luminosity was estimated by Equation (4). Finally, the lasting time of the jets was calculated, which cannot break out from the envelope. The main parameters are given as $f=0.9$, $\eta=10^{-5}$, and $\theta_{\rm j}=10^\circ$. The BH mass value is not sensitive to the results, which can be reasonably set as $\sim 5~M_\odot$. The mean SN luminosity in the whole accretion timescale, $\sim 10^{42}~\rm erg ~s^{-1}$, is satisfied with iPTF14hls. The symbols s, o, and z represent the metallicity values of the progenitors $Z=Z_\odot$, $10^{-1}~Z_\odot$, and 0, respectively, where $Z_\odot$ is the metallicity of the Sun. The numbers 20, 30, 40, 50, 60, 75, and 80 represent the masses of the progenitor stars in units of solar mass.
	
The data of Figure 2 are calculated by using the density profiles of the pre-SN model \citep[e.g.,][]{Heger2003,Woosley2012}. The processes of the SN explosions were not included in our calculations. Here, we only show that in the whole accretion process, which lasts approximately 300 days (roughly equal to the free-fall timescale), the jets cannot break out from the envelope. In addition, the accretion timescale of other types of stars with $10^{-2}~Z_\odot$ and $10^{-4}~Z_\odot$ cannot last more than approximately 200 days, which are not shown in this figure.
	
In the pre-SN model, the accretion rate is roughly constant after tens of seconds \citep[e.g.,][]{Liu2018}. Once the outflow feedback mechanism is considered, the mass supply rate is relatively unaffected by the outflows mentioned above. Thus, one can expect that, for the progenitor stars as shown in Figure 2, if the strong outflows with $f=0.9$ inject the envelope in the endless cycle and the outflows are assumed to never break out from the envelope, the accretion timescale can at best last nearly 10 times (the sum of an infinite geometric progression with the multiple of 0.9) that of the free-fall timescale. Thus, accretion process can be sustained for no more than 3,000 days. The massive progenitors in Figure 2 might be adequate for iPTF14hls to last at least 1,153 days \citep{Andrews2018} or even over a much longer timescale \citep{Chugai2018,Sollerman2018}. In brief, we infer that the progenitor of iPTF14hls is a low-matellicity and massive star in the disk outflow feedback scenario, although there is no observable evidence for now.
	
In addition, the lower $\eta$ and  larger $\theta_{\rm j}$ are satisfied with the choked jets and the luminosity of bright SNe for the longer accretion timescale once including the outflow feedback effects. Meanwhile, in addition to the contribution of the choked jet, the total luminosity of SNe might include the decay of some radioactive elements (such as $^{56}$Ni and $^{56}$Co) produced by the outflows \citep[e.g.,][]{Surman2006,Song2019}. The decreased inflow rate causes the luminosity to decrease. The choked jet sustains a loss of dominance on the luminosity. Once these elements are almost exhausted, the luminosity may quickly decrease.
	
\subsection{Accretion rate fluctuations}
	
Another challenge of explaining iPTF14hls based on the BH hyperaccretion process is the existence of at least five peaks in its light curve.
	
In our case, if we still assume that the mass and energy of the outflows are isotropically injected into the envelope, we can estimate the length of time needed to convert the outflows into the fall materials by the virial theorem, i.e.,
\beq
\int_{\Delta t} \dot{M}_{\rm o} v^2_{\textrm{o}}  d t \approx \int_{>R} \frac{G M (<R)d M_R}{R},
\eeq
where $R$ is the radius of the star.
	
This means that the kinetic energy accumulations of the outflows can at least partly prevent the falling process of materials from the envelope, thereby modulating firstly the mass apply rate and then the inflow and outflow rates. For $\dot{M} \sim 10^{-5}~M_\odot~\rm s^{-1}$ (corresponding to the typical accretion rate of most of the progenitors in Figure 2), $v_{\textrm{o}}=0.1~c$, and $f=0.9$, the timescale $\Delta t$ is approximately 100 days for the massive progenitors, $\sim 50~M_\odot$, which coincide with the timescale of the peaks of iPTF14hls. Here, the right-hand side of Equation (9) can be integrated by using the density profiles of the progenitor stars \citep[e.g.,][]{Liu2018}. The fluctuation of the mass application rate leads that of the inflow rate. The inflow rate determines the jet luminosity; then, the continual multi-peak light curve is produced. However, the outflows are indeed in the anisotropic distribution, in the continual modulations and games between the outflows and the supply mass of the envelope, the fluctuation of the mass supply rate should not be violent but be mild.
	
It is well known that many SNe have a double-peak light curve \citep[e.g.,][]{Arnett1989,Mazzali2008,Wang2018}. Once the feature cannot be explained by the shock cooling mechanism \citep[e.g.,][]{Piro2015}, the outflow feedback might be worth taking into account.
	
\section{Conclusions and discussion}
	
The outflow feedback from the BH hyperaccretion system in the collapsar displays the following two types of effects: prolonging the accretion timescale and controlling the varieties of the accretion rate. Jet alignment along the line of sight cannot break out from the envelope of a massive progenitor star or its CSM might power iPTF14hls. The unusual characteristics of this jet-driven SN might be vividly represented by the influences of the strong outflows from the disk in a massive progenitor star. From the above estimations, iPTF14hls can be expected to survive for no longer than approximately 3,000 days, with the luminosity decreasing quickly in the late phase \citep{Sollerman2018}.
	
The disk outflows can be driven by the radiation pressure or the large-scale magnetic fields rooted in the accretion disk. In the case of the powerful disk outflows in the collapsars, the strong magnetic fields are the plausible origins, which should require the progenitors to be magnetized \citep[e.g.,][]{Takiwaki2009}.
	
In this paper, we only estimate the disk outflow feedback for iPTF14hls. More detailed and time-dependent descriptions of the BH inflow-outflow hyperaccretion systems in the collapsar scenario should be performed, including the global mass and velocity distributions of the outflows, the global mass distribution of the envelope continuously colliding with the outflows, the time delay of the change transmission among the mass apply rate, inflow rate, and outflow rate, and the modulation of the angular momenta amongst the progenitor, inflows and outflows. Moreover, the mechanisms to prolong the accretion timescale or fluctuate the accretion rate, such as the value or evolution of the viscosity and the magnetic barrier \citep[e.g.,][]{Proga2006,Liu2012}, might be considered. In the future, we will further simulate the outflow feedback model in the LGRB-SN scenario.
	
In addition, vertical advection (or convection) possibly occurs in the super-Eddington accretion disks \citep[e.g.,][]{Jiang2014,Liu2015,Liu2017,Yi2017}. The gamma-ray photons produced in the hyperaccretion disk will be trapped in the bubbles. The magnetic buoyancy forces the bubbles to rise to the disk surface, and the photons escape. Gamma-ray luminosity above $10^{50}~\rm erg~s^{-1}$ can be achieved when the accretion rate is larger than $10^{-3}~M_\odot~\rm s^{-1}$ \citep{Yi2017}, so gamma-ray radiation feedback may exist in collapsars.
	
\section*{Acknowledgments}
We thank Prof. Alexander Heger for supplying the pre-SN data. This work was supported by the National Natural Science Foundation of China under Grant Nos. 11822304 and 11333004. X Wang is supported by the National Natural Science Foundation of China under Grant Nos. 11325313 and 11633002.

\end{document}